\documentclass{PoS}

\title{$\mathcal N=3$ four dimensional field theories}

\ShortTitle{$\mathcal N=3$ four dimensional field theories}

\author{I\~naki Garc\'ia-Etxebarria\\
        Max Planck Institute for Physics, F\"ohringer Ring 6, 80805 Munich, Germany\\
        E-mail: \email{inaki@mpp.mpg.de}}

\author{\speaker{Diego Regalado}%
\\
       Max Planck Institute for Physics, F\"ohringer Ring 6, 80805 Munich, Germany\vspace{0.2cm}\\
       Institute for Theoretical Physics and Center for Extreme Matter and Emergent Phenomena,
       Utrecht University, Leuvenlaan 4, 3584 CE Utrecht, The Netherlands\\
       E-mail: \email{regalado@mpp.mpg.de}}

\abstract{We briefly review a class of four dimensional $\mathcal N=3$ field theories constructed by taking a quotient of $\mathcal N=4$ SYM with gauge group $U(N)$. The quotient involves a discrete symmetry that only exists for specific, order one, values of the coupling constant, so the resulting theories are intrinsically strongly coupled. These theories admit a simple realization in string theory as the worldvolume theory of a stack of D3 branes probing a generalized orientifold plane, or S-fold. Their holographic dual is given by a non-trivial F-theory fibration over $AdS_5 \times S^5/\mathbb Z_k$ which is weakly curved but with the string coupling frozen at an order one value.}

\FullConference{Corfu Summer Institute 2016 "School and Workshops on Elementary Particle Physics and Gravity"\\
                 31 August - 23 September, 2016\\
                 Corfu, Greece}

\begin{document}

\section{Introduction and summary}

Four dimensional $\mathcal N=3$ field theories have been largely ignored in the past since a well-known argument shows that Lagrangians with $\mathcal N=3$ supersymmetry are actually $\mathcal N=4$ SYM.\footnote{The $\mathcal N=3$ gravity multiplet is different from the $\mathcal N=4$ multiplet so genuine supergravity Lagrangians with $\mathcal N=3$ supersymmetry in four dimensions can be constructed.} However, such argument is based on the existence of a Lagrangian which is not a necessary requirement to define a consistent quantum field theory. Thus, non-Lagrangian $\mathcal N=3$ field theories are not ruled out.

The possibility of having $\mathcal N=3$ SCFTs, which do not enhance to $\mathcal N=4$, was recently considered in \cite{Aharony:2015oyb} (see also \cite{Cordova:2016xhm,Cordova:2016emh}) where many properties of such theories were obtained by analyzing the superconformal algebra, under the assumption that they exist. This analysis shows, in particular, that the conformal anomalies of these theories satisfy $a=c$ and that they cannot have any continuous global symmetries beyond the R-symmetry $SU(3)_R\times U(1)_R$. In this sense, $\mathcal N=3$ SCFTs are similar to those with $\mathcal N=4$ supersymmetry. However, a crucial difference between them is that $\mathcal N=3$ SCFTs do not have $\mathcal N=3$-preserving marginal deformations so they are isolated fixed points. This may explain why they have not been discussed in the past.

The purpose of this proceedings article is to provide a short review of the simplest known examples of four dimensional non-Lagrangian $\mathcal N=3$ theories and their string theory realization \cite{Garcia-Etxebarria:2015wns}. These are constructed by taking certain quotient of $\mathcal N=4$ SYM with gauge group $U(N)$ by a non-perturbative discrete symmetry present only for particular values of the coupling constant (see \cite{Ferrara:1998zt} for an early discussion of this possibility from a holographic perspective). The quotient projects out four of the original sixteen supercharges together with the marginal deformation, in agreement with the general results of \cite{Aharony:2015oyb,Cordova:2016xhm}.

These theories appear naturally in string theory as the worldvolume theory of a stack of D3 branes probing an S-fold, which is a generalization of the O3 plane involving non-trivial elements of the S-duality group of Type IIB.\footnote{As we will see, from this viewpoint the O3 plane is the special case in which the S-duality element is $(-1)\in SL(2,\mathbb Z)$.} Alternatively, they can be constructed by taking the F-theory limit of a stack of M2 branes probing an orbifold of the form $(\mathbb C^3\times T^2)/\mathbb Z_k$. This geometric picture shows that these $\mathcal N=3$ theories flow to ABJM theories \cite{Aharony:2008ug,Aharony:2008gk} when compactified on a circle. The holographic dual is similarly obtained by taking a quotient of the well-known Type IIB $AdS_5\times S^5$ which fixes the string coupling constant to an order one value.

\section{Field theory construction}

Let us start by considering four dimensional $\mathcal N=4$ SYM with gauge group $U(N)$ and coupling constant $\tau$. As is well known, this theory has an R-symmetry group $SU(4)_R$ under which the supercharges transform in the antifundamental representation. It is also self-dual under Montonen-Olive duality \cite{Montonen:1977sn}, in the sense that the theory with coupling constant $\tau$ is mapped to another theory with the same gauge group and with coupling constant $\tau'$ given by
\begin{equation}\label{duality}
\tau'=\frac{a\tau+b}{c\tau+d}\qquad \qquad \left( \begin{array}{cc}a&b\\c&d\end{array}\right )\in SL(2,\mathbb Z)\,.
\end{equation}
It is important to stress that the duality group $SL(2,\mathbb Z)$ is not a symmetry of the theory, since generically it maps a theory with coupling constant $\tau$ to a \emph{different} theory with coupling constant $\tau'\neq \tau$. However, for particular values of $\tau$, there can be a subgroup $\Gamma$ of $SL(2,\mathbb Z)$ that leaves $\tau$ invariant, in which case the discrete group $\Gamma$ is indeed a global symmetry of the theory defined at that particular coupling.  Since the element $(-1)\in SL(2,\mathbb Z)$ does not act on $\tau$, we find that there is a discrete symmetry $\mathbb Z_2$ for any $\tau$. In addition to that, there are two special values of $\tau$ for which there is a further enhancement of the symmetry, as shown in table \ref{tab:1}.

\begin{table}
  \centering
  \begin{tabular}{c|c|c}
    $\tau$ & $\Gamma$ & Generator \\
    \hline
    any & $\mathbb Z_2$ & $\left( \begin{array}{cc}-1&0\\0&-1\end{array}\right )$ \\
    $i$ & $\mathbb  Z_4$ & $\left( \begin{array}{cc}0&-1\\1&0\end{array}\right )$ \\
    $e^{i \pi/3}$ & $\mathbb  Z_6=\mathbb Z_2\times \mathbb Z_3$ & $\left( \begin{array}{cc}1&-1\\1&0\end{array}\right )$ \\
  \end{tabular}
  \caption{Discrete symmetries of $\mathcal N=4$ SYM with gauge group $U(N)$ and coupling constant $\tau$.}
  \label{tab:1}
\end{table}

Once we have identified the values of $\tau$ for which there is a discrete S-duality symmetry $\mathbb Z^{S}_k$, we can consider the theory that results after taking the quotient by such a discrete symmetry.\footnote{We will clarify what we mean by quotient in the next section when we discuss the string theory embedding of this construction.} The supercharges transform non-trivially under a duality transformation \label{duality} as \cite{Kapustin:2006pk}
\begin{equation}
Q_{\alpha a}\rightarrow \left (\frac{c\tau+d}{|c\tau+d|} \right )^{\frac{1}{2}} Q_{\alpha a}\,,
\end{equation}
so under the $\mathbb Z^S_k$ subgroup of $SL(2,\mathbb Z)$ they pick up a phase
\begin{equation}
Q_{\alpha a}\rightarrow e^{i\pi/k} Q_{\alpha a}\,.
\end{equation}
Here $a$ is an index in the antifundamental of $SU(4)_R$ which runs from 1 to 4.
This means that the quotient by $\mathbb Z_k^S$ breaks supersymmetry completely since none of the supercharges are invariant. However, we may combine the S-duality symmetry $\mathbb Z^S_k$ with a $\mathbb Z_k$ subgroup of the R-symmetry group in order to preserve as much supersymmetry as possible. Under the subgroup $\mathbb Z^R_k$ generated by (as an element of $SO(6)_R$)
\begin{equation}
R_k=\left( \begin{array}{ccc}
\hat R_k&0&0\\
0&\hat R_k^{-1}&0\\
0&0&\hat R_k\end{array}\right )\,,
\end{equation}
where $\hat R_k$ denotes a rotation by $2\pi/k$ on $\mathbb R^2$, the supercharges transform as
\begin{equation}
Q_{\alpha A}\rightarrow e^{-i\pi/k} Q_{\alpha A} \qquad \qquad Q_{\alpha 4}\rightarrow e^{3i\pi/k} Q_{\alpha 4}
\end{equation}
where $A=1,2,3$. Thus, under the combined action
\begin{equation}\label{qq}
\mathbb Z_k=\mathbb Z^S_k\cdot \mathbb Z_k^R
\end{equation}
we have that
\begin{equation}
Q_{\alpha A}\rightarrow Q_{\alpha A} \qquad \qquad Q_{\alpha 4}\rightarrow e^{4i\pi/k} Q_{\alpha 4}\,.
\end{equation}
We find that the quotient (\ref{qq}) preserves sixteen supercharges for $k=2$ but only twelve survive for $k=3,4,6$, which corresponds to $\mathcal N=3$ supersymmetry in four dimensions.\footnote{This kind of quotient was previously discussed in \cite{Ferrara:1998zt} (see also \cite{Ganor:2010md} for a similar construction in the case of compactification on a circle).} Notice that the values of $k$ for which supersymmetry gets reduced are precisely those that exist only for a particular value of the coupling constant $\tau$. This is in agreement with the analysis carried out in \cite{Aharony:2015oyb,Cordova:2016xhm} which shows that genuine $\mathcal N=3$ SCFTs do not have marginal deformations preserving all the supersymmetry. 

Notice that even thought our starting point is four dimensional $\mathcal N=4$ SYM with gauge group $U(N)$, the whole discussion can be extended to the ADE-type groups, as done in \cite{Garcia-Etxebarria:2016erx}. 

In the following we provide an explicit string theory realization of the $\mathcal N=3$ theories of type A, which is useful to analyze some of their most basic properties. For an M-theoretic construction of the theories of exceptional type, see \cite{Garcia-Etxebarria:2016erx}. We should also mention that a complementary approach using bootstrap techniques has been recently initiated in \cite{Lemos:2016xke,Cornagliotto:2017dup}. Furthermore, the $\mathcal N=3$ theories described here fit nicely in the classification of rank-one $\mathcal N=2$ SCFTs of \cite{Argyres:2015ffa,Argyres:2015gha,Argyres:2016xua,Argyres:2016xmc,Argyres:2016yzz}.

\section{String theory embedding}


Consider $N$ M2 branes on $\underline{\mathbb R^{1,2}}\times \mathbb C^3\times T^2$, where the underlining denotes the dimensions wrapped by the M2 branes. At low energies, the theory living on the M2 branes is the well-known ABJM theory at level one \cite{Aharony:2008ug}. Since we are interested in having a four dimensional theory, we may take the so-called F-theory limit \cite{Vafa:1996xn}, under which the M2 branes become D3 branes, whose worldvolume is described by $\mathcal N=4$ SYM with gauge group $U(N)$ and with coupling constant given by the complex structure $\tau$ of the F-theory torus.\footnote{The F-theory limit consists of three steps: first reduce along one of the circles in the torus which gives Type IIA with $N$ D2 branes on $\underline{\mathbb R^{1,2}}\times \mathbb C^3\times S_T^1$, second T-dualize along the remaining circle arriving at Type IIB with $N$ D3 branes on $\underline{\mathbb R^{1,2}\times \tilde S_T^1}\times \mathbb C^3$. Finally, taking the size of $S^1_T$ to zero, we end up with $N$ D3 branes on $\underline{\mathbb R^{1,3}}\times \mathbb C^3$. The axio-dilaton $\tau_{IIB}$ is given by the complex structure of the original torus which is then the coupling constant of the SYM theory living on the stack of D3 branes.} The advantage of starting with the description in terms of M2 branes instead of considering directly D3 branes is that in the former, both the R-symmetry group of the four dimensional theory and the Montonten-Olive duality group are manifest geometrically. Indeed, the R-symmetry group $SO(6)_R$ corresponds to the rotations in $\mathbb C^3$ leaving the origin fixed and the $SL(2,\mathbb Z)$ duality group comes from the large diffeomorphisms of the F-theory torus, which is geometric in three dimensions.

Given this realization of four dimensional $\mathcal N=4$ SYM with gauge group $U(N)$ it is straightforward to implement the quotient introduced in the previous section. We can just consider $N$ M2 branes\footnote{Here we consider $N$ to be the number of mobile M2 branes, that is, the rank of the resulting theory.} on the orbifold $\underline{\mathbb R^{1,2}}\times (\mathbb C^3\times T^2)/\mathbb Z_k$ defined by
\begin{equation}\label{asd}
(z^1,z^2,z^3,u)\rightarrow (\zeta_k z^1, \bar \zeta_k z^2, \zeta_k z^3, \bar \zeta_k u)\,,\qquad \qquad \zeta_k=e^{2\pi i /k}\,,
\end{equation}
where $(z^1,z^2,z^3,u)$ are complex coordinates on $\mathbb C^3\times T^2$. From this perspective we recover the fact that the quotient is only well defined for the values of $\tau$ and $k$ that we found in field theory (see table \ref{tab:1}), as well as the amount of supersymmetry preserved by this M-theory configuration. Indeed, the ABJM construction preserves sixteen supercharges for $k=1,2$ and twelve for $k>2$.

\medskip

In order to obtain the four dimensional theory from the M-theory setup with M2 branes we have to take the F-theory limit, as mentioned earlier. The cases in which we preserve only twelve supercharges correspond to Type IIB setups in which the axion-dilaton (which contains the string coupling constant) is frozen to an order one value, which makes the analysis difficult. However, the case in which $k=2$ turns out to be very illuminating since the coupling constant $\tau$ is not projected out so it can be analyzed in perturbation theory. As shown in \cite{Hanany:2000fq}, upon taking the F-theory limit of the orbifold defined by (\ref{asd}) for $k=2$, we end up in Type IIB in the presence of an ${\rm O3}^{-}$ plane. Indeed, an O3 plane is defined as Type IIB on $\mathbb R^{1,3}\times \mathbb C^3/(\mathcal I\cdot \Omega\cdot (-1)^{F_L})$ where $\mathcal I$ acts geometrically on $\mathbb C^3$ by inverting the coordinates, $\Omega$ is orientation reversal on the worldsheet and $(-1)^{F_L}$ is the left moving spacetime fermion number operator. The F-theory description of the action on the worldsheet $\Omega\cdot (-1)^{F_L}$ is simply an inversion of the F-theory torus, as can be checked by looking at its action on the Type IIB massless fields. The combined action $\mathcal I\cdot \Omega\cdot (-1)^{F_L}$ is therefore given by the orbifold (\ref{asd}) for $k=2$. This allows us to conclude that for $k=2$ the theory living on the worldvolume of the probe D3 branes is four dimensional $\mathcal N=4$ SYM with gauge group $O(2N)$, as can be obtained from open string perturbation theory. Notice that the parent theory in this case is SYM with gauge group $U(2N)$ so we see that the quotient we are considering is not a gauging in spacetime since it changes drastically the local physics.\footnote{The theory with gauge group $O(2N)$ can be obtained from the $SO(2N)$ theory by a discrete gauging of the $\mathbb Z_2$ global symmetry of the $SO(2N)$ theory.}

There is more that can be learned from the perturbative case $k=2$. As is well known \cite{Witten:1998xy}, there is not just one type of O3 plane but there are four of them, usually denoted by ${\rm O3}^{-},{\rm O3}^{+},\widetilde {\rm O3}^{-},\widetilde {\rm O3}^{+}$. The first one is a singlet under the Type IIB  $SL(2,\mathbb Z)$ duality and the other three form a triplet. The theory of $N$ D3 branes probing the ${\rm O3}^{+}$ is $\mathcal N=4$ SYM with gauge group $USp(2N)$ and since the other cases are dual to this one, they correspond to the same theory at a different point in the conformal manifold.\footnote{From the F-theory perspective, the different variants are encoded in the discrete flux that can be turned on at the orbifold singularities \cite{Hanany:2000fq}. Some choices of discrete fluxes in three dimensions correspond to shift-orientifolds so they do not give rise to four dimensional orientifolds.} Thus, for these variants the four dimensional theory compactified on a circle may flow to the ABJ theories with fractional M2 branes \cite{Aharony:2008gk} (see also \cite{Gang:2011xp}). The conclusion is that the four dimensional quotient is not uniquely determined by specifying the value of $k$, since for $k=2$ there are two different cases. In addition to that, from a purely field theory perspective, one could consider discrete gaugings which give further possibilities. However, these discrete gaugings are relatively mild modifications of the theory that do not change the local dynamics, even thought they change the spectrum of local and non-local operators. For a discussion of discrete gaugings in the context of $\mathcal N=3$ theories, see \cite{Aharony:2016kai,Argyres:2016yzz}.

The cases with $\mathcal N=3$ supersymmetry, namely $k=3,4,6$, correspond to the worldvolume theory of D3 branes probing generalized orientifold planes usually referred to as S-folds, since the orientifold action involves a non-trivial `S-duality orbifold'. In these cases, the marginal deformation is projected out so we cannot rely on open string perturbation theory to analyze the dynamics on the probe D3 branes. However, the three dimensional theory obtained by compactifying on a circle is quite similar in both cases, namely ABJ(M) theories. In particular, this suggests that there may be different variants for a fixed value of $k$, which were classified in \cite{Aharony:2016kai,Imamura:2016abe}. The central charges of these theories were obtained in \cite{Aharony:2016kai} by combining the fact that $a=c$ for $\mathcal N=3$ SCFTs \cite{Aharony:2015oyb} and the Shapere-Tachikawa relation between the central charges of $\mathcal N=2$ SCFTs and the dimension of the Coulomb branch operators \cite{Shapere:2008zf}. 

The case of rank-one, corresponding to a single mobile D3 brane, was considered in \cite{Nishinaka:2016hbw} where it was shown that the moduli space allowed by superconformal invariance agrees nicely with the one expected from the motion of a single D3 brane in the presence of an S-fold, namely $\mathbb C^3/\mathbb Z_k$ for $k=3,4,6$. The chiral algebra, in the sense of \cite{Beem:2013sza}, has been derived for the rank-one case in \cite{Nishinaka:2016hbw} and for the higher-rank cases it has been proposed in \cite{Lemos:2016xke}. Further aspects of these theories have been studied in \cite{Imamura:2016udl,Agarwal:2016rvx}.

\subsection{Holographic dual}

It is possible to obtain the holographic dual description of the $\mathcal N=3$ theories by taking a quotient of the usual Type IIB $AdS_5\times S^5$ solution, dual to a large number of D3 branes. Again, it is useful to start by considering the case $k=2$, namely the holographic dual of D3 branes probing an O3 plane \cite{Witten:1998xy}. The inversion of the coordinates transverse to the O3 plane $\mathcal I$ maps to a $\mathbb Z_2$ involution acting on $S^5$, so the internal space becomes $\mathbb{RP}^5=S^5/\mathbb Z_2$. In addition to that, we have to include the action on the worldsheet $\Omega\cdot (-1)^{F_L} $, which can be thought of as an $SL(2,\mathbb Z)$ bundle on $\mathbb{RP}^5$ with transition functions in $\mathbb Z_2$. This background can be regarded as F-theory on $AdS_5\times (S^5\times T^2)/\mathbb Z_2$, where the $\mathbb Z_2$ is acting on both $S^5$ and $T^2$, so that the resulting space is not simply a product.

From the holographic perspective, the different types of O3 planes are characterized by the discrete fluxes on $\mathbb{RP}^5$ \cite{Witten:1998xy}. The Type IIB two-forms $B_{NS}$ and $B_{RR}$ are odd under the action of $\Omega\cdot (-1)^{F_L}$, which means that the associated fluxes on $\mathbb{RP}^5$ are classified by $H^3(\mathbb{RP}^5,\widetilde{\mathbb Z})=\mathbb Z_2$, where the tilde indicates that these are sections of the bundle of three-forms twisted by the action of $\Omega\cdot (-1)^{F_L}$. Therefore, this gives four different possibilities for the O3 plane, depending on the choice of discrete fluxes. The one without fluxes is invariant under the duality group and corresponds to the ${\rm O3}^-$. The other three choices, which are transformed into each other by the duality group, give rise to the triplet ${\rm O3}^+,\widetilde{\rm O3}^-,\widetilde{\rm O3}^+$. 

The holographic dual of the genuine $\mathcal N=3$ theories can be obtained similarly. In these cases, the internal space is given by $S^5/\mathbb Z_k$ and there is an S-duality bundle with transition functions on $\mathbb Z_k$ acting as we go around the non-trivial one-cycle of $S^5/\mathbb Z_k$. Alternatively, we may consider an F-theory compactification on $AdS_5\times (S^5\times T^2)/\mathbb Z_k$. Notice that $\mathbb Z_k$ acts freely on the sphere $S^5$ so the resulting Type IIB geometry is smooth and weakly curved for a large number of D3 branes.\footnote{It is due to the non-trivial global structure that this construction evades the no-go result of \cite{Beck:2016lwk}.} However, for $k>2$ the quotient freezes the string coupling constant to an order one value, so both the SCFT and holographic dual are necessarily strongly coupled. Notice that the classification of the perturbative O3 plane variants does not rely on worldsheet techniques, so it can be applied to classify these non-perturbative $\mathcal N=3$ S-folds \cite{Aharony:2016kai,Imamura:2016abe}. Such analysis shows that there are two different S-fold variants for $k=3,4$ and only one for $k=6$. 

The holographic dual can be used to compute the superconformal index of the $\mathcal N=3$ theories in the large $N$ limit \cite{Imamura:2016abe}. The leading contribution comes from the Kaluza-Klein modes in $AdS_5\times S^5/\mathbb Z_k$ together with the action of $SL(2,\mathbb Z)$, that are obtained by truncating those of $AdS_5\times S^5$. Finite $N$ corrections, which distinguish between the different variants for a fixed values of $k$, have also been studied \cite{Imamura:2016abe}.

\medskip
\medskip

\emph{Acknowledgments}: This article summarizes a talk given at the "Workshop on Geometry and
Physics", which took place on November 20-25, 2016 at the Ringberg Castle, Germany. The workshop
was dedicated to the memory of Ioannis Bakas.

\end{document}